# Evaluation of PCA Image Denoising on Multi-Exponential MRI Relaxometry

**Mark D. Does PhD**[1,2,3,4] | **Jonas Lynge Olesen BSc**[5,6] | **Kevin D. Harkins PhD**[1,2] | **Teresa Serradas-Duarte MSc**[7] | **Daniel F. Gochberg PhD**[2,3] | **Sune N. Jespersen PhD**[5,6] | **Noam Shemesh PhD**[7]

[1]Department of Biomedical Engineering, Vanderbilt University, Nashville, TN, US

[2]Institute of Imaging Science, Vanderbilt University Medical Center, Nashville, TN, US

[3]Department of Radiology and Radiological Sciences, Vanderbilt University Medical Center, Nashville, TN, US

[4]Department of Electrical Engineering, Vanderbilt University, Nashville, TN, USA

[5]Center of Functionally Integrative Neuroscience, Aarhus University Hospital, Aarhus, Denmark

[6]Department of Physics and Astronomy, Aarhus University, Aarhus, Denmark

[7]Champalimaud Centre for the Unknown, Lisbon, Portugal

**Correspondence**
Mark D. Does
Email: mark.does@vanderbilt.edu

**Funding information**
European Research Council, 679058; The Danish National Research Foundation the Danish Ministry of Science, Innovation, and Education; The National Institutes of Health, EB019980

**Purpose**: Multi-exponential relaxometry is a powerful tool for characterizing tissue, but generally requires high image signal-to-noise ratio (SNR). This work evaluates the use of principal-component-analysis (PCA) denoising to mitigate these SNR demands and improve the precision of relaxometry measures.

**Methods**: PCA denoising was evaluated using both simulated and experimental MRI data. Bi-exponential transverse relaxation signals were simulated for a wide range of acquisition and sample parameters, and experimental data were acquired from three excised and fixed mouse brain. In both cases, standard relaxometry analysis was performed on both original and denoised image data, and resulting estimated signal parameters were compared.

**Results**: Denoising reduced the root-mean-square-error of parameters estimated from multi-exponential relaxometry by factors of 2 to 4×, depending on the acquisition and sample parameters. Denoised images and subsequent parameter maps showed little or no signs of spatial artifact or loss of resolution.

**Conclusion**: Experimental studies and simulations demonstrate that PCA denoising of MRI relaxometry data is an ef-fective method of improving parameter precision without sacrificing image resolution. This simple yet important processing step thus paves the way for broader applicability of multi-exponential MRI relaxometry.

# 1 | INTRODUCTION

Multi-exponential MRI relaxometry is a powerful tool for characterizing tissue at the sub-voxel level, the most well known example of which is the use of multi-exponential $T_2$ (MET$_2$) relaxometry for measuring myelin content in brain [1, 2] and nerve [3]. Similar myelin imaging has been implemented based on multi-exponential $T_2^*$ relaxometry [4, 5], and other example applications of multi-exponential MRI relaxometry include characterization of muscle [6, 7, 8, 9], cartilage [10, 11], and tumors [12, 13]. Similarly, inversion-recovery based quantitative magnetization transfer is effectively a bi-exponential signal analysis [14]. However, the challenge in making effective use of MET$_2$ or other multi-exponential relaxometry stems from the ill-posed nature of inverting a signal into a linear combination of exponential functions [15]. For many situations, including myelin water imaging, image signal-to-noise ratio (SNR) in the 100s is required for reasonable results [16, 17]. Such high SNR values dictate low resolution and/or long scan times, which has limited the use of multi-exponential MRI relaxometry.

These SNR demands can be mitigated to some extent through conventional image-domain filtering [18] or joint regularization in the spectral and spatial domains [19, 20], although there is a corresponding cost in spatial resolution of the fitted parameter maps. Edge-preserving image-domain filters have been used (for example, anisotropic diffusion [21] and non local means [22] filters), but the fundamental trade-off between parameter map precision and resolution remains. Recent work using low rank representations of MRI signals offers an alternative approach to improving parameter maps precision. For example, the redundancy of information in multiple spin echo images has been used to produce higher quality images [23] and faster acquisition for mono-exponential $T_2$ mapping [24, 25, 26]. Further, low rank denoising has been demonstrated effective for improving precision of chemical shift imaging data [27], and fat fraction maps [28, 29] derived from multiple gradient echo acquisitions. Similarly, principal component analysis (PCA) has been used for denoising diffusion weighted imaging (DWI) data sets for improved precision of diffusion tensor or other parametric characterizations [30, 31, 32]. To our knowledge, no such low rank approach has been evaluated for improving precision of parameter maps derived from multi-exponential signal characterizations. Thus, we present an evaluation of a particular PCA denoising method [32] for multi-exponential MRI relaxometry data, with the specific aim of improving precision of myelin maps derived from MET$_2$ or qMT MRI methods.

# 2 | METHODS

## 2.1 | Simulations

Relaxometry data were simulated to represent human brain data at $3.0$ T. Model multiple spin echo images, $I$, of dimension $N_x \times N_y = 100 \times 100$ were generated at spin echo times $t = T_E$ to $N_E T_E$. The signal in each voxel was defined as the sum of two exponential decays,

$$I(j, k, m) = f_s(j, k) \exp\left[\frac{-t(m)}{T_{2s}(j, k)}\right] + (1 - f_s(j, k)) \exp\left[\frac{-t(m)}{T_{2l}(j, k)}\right],$$
$$j = 1 \text{ to } N_x, \ k = 1 \text{ to } N_y, \ m = 1 \text{ to } N_E. \tag{1}$$

In all cases, $T_{2s} < T_{2l}$, and so the parameter map, $f_s$, was the map of the short $T_2$ signal fraction. This map was constructed with $f_s$ values ranging 0 to 0.25 in vertical strips of non-zero value with widths ranging 1 to 9 pixels (see Fig 1). Within these strips, the simulated signals were meant to reflect white matter signals, with $f_s$ being the myelin water fraction (MWF), and between the strips, where $f_s = 0$, the simulated signals were meant to reflect gray matter signals. For each pixel location, $(j, k)$, the relaxation time constants, $T_{2s}$ and $T_{2l}$, were randomly drawn from normal distributions, $\mathcal{N}(\mu_{T2s} = 15\,\text{ms}, \sigma_{T2s} = 1\,\text{ms})$ and $\mathcal{N}(\mu_{T2l} = 80\,\text{ms}, \sigma_{T2l})$, respectively. Simulations were repeated for $\sigma_{T2l} = 1$ to $10\,\text{ms}$, in $1\,\text{ms}$ steps, and for 5 $N_E, T_E$ combinations spanning $26{,}12\,\text{ms}$ to $80{,}4\,\text{ms}$. For each combination of signal parameters, a complex noisy image, $I_n$, was generated as

$$I_n(j, k, m) = I(j, k, m) + \eta_{re}(j, k, m) + i\eta_{im}(j, k, m) \,, \tag{2}$$

where, values $\eta_{re}(i, j, k)$ and $\eta_{im}(i, j, k)$ were each randomly drawn from a normal distribution, $\mathcal{N}(0, \sigma)$. Simulations were repeated across a log-space range of SNR values, with SNR $\triangleq 1/\sigma = 25$ to $1600$.

## 2.2 | MRI

Animal experiments were approved by the Vanderbilt Institutional Animal Care and User Committee. In order to test the efficacy of the denoising on experimental data, chemically fixed mouse brains were imaged using a 7 T 16 cm horizontal bore Bruker BioSpec scanner (Rheinstetten, Germany), using a 25 mm diameter quadrature volume coil for transmission and reception (Doty Scientific, Columbia, SC). The brains of 3 adult mice were perfusion-fixed (2.5% glutaraldehyde + 2% paraformaldehyde), washed of fixative, and loaded (during and following perfusion) with 1.0 mM gadolinium (Prohance; Bracco Diagnostics, Princeton, NJ). To provide a signal-free background and prevent tissue dehydration, brains were placed in an MR-compatible tube filled with perfluoropolyether liquid (Fomblin, Solvay Solexis, Thorofarem NJ, USA). For each brain, a single 0.75 mm thick coronal slice was selected to provide relatively uniform white matter through plane in the middle of the corpus callosum. Two quantitative imaging protocols were run, each with with $150 \times 150\,\mu\text{m}^2$ in plane resolution: i) multiple spin echo (MSE) with refocusing to excitation bandwidth ratio = 4.6, $N_E = 96$, $T_E = 5\,\text{ms}$ (uniform echo spacing), $T_R = 800\,\text{ms}$, and number of averaged excitations ($N_A$) = 4; and ii) inversion-recovery (IR) prepared RARE with 5 ms echo spacing, centric phase encoding, $N_I = 15$ inversion times log-spaced from 5 to 1500 ms, and $N_A = 1$. For each brain, each protocol was repeated 32 to 64 times and the complex image data were cumulatively averaged. This resulted in images sets with $N_A$ up to 256 for the MSE scans up to 64 for the IR scans.

## 2.3 | Data analysis

### 2.3.1 | Denoising

The PCA image denoising algorithm used in this study was presented in detail by Veraart et al. [32] and is briefly summarized here in the context of relaxometry. Let $\mathbf{X} \in \mathbb{C}^{N_q \times N_v}$ be signals from a subset of $N_v$ voxels acquired with $N_q$ different signal contrasts. In the context of the studies presented here, $N_q = N_E$ for the multiple spin echo images or $N_q = N_I$ for the inversion-recovery prepared images. Let $\bar{\mathbf{x}} \in \mathbb{C}^{N_q \times 1}$ be the mean signal across the $N_v$ voxels, and then the mean-removed signal matrix can be factored by singular value decomposition, $\left(\mathbf{X} - \bar{\mathbf{x}}\mathbf{1}_{1 \times N_v}\right) = \mathbf{U}\mathbf{S}\mathbf{V}^\mathsf{T}$. From here, the eigenvalues of the sample data covariance matrix are, $\lambda_i = S_{i,i}^2/N_v$, $i = 1$ to $M$, where $M = \min(N_v, N_q)$. For a sufficient SNR, one can assume that the $P$ largest eigenvalues correspond to the eigenvectors that predominantly characterize signal, and the remaining $M - P$ eigenvectors predominantly characterize the added noise. For independent and identically distributed (iid) zero-mean random noise, $\lambda$ is a random variable described by the Marcenko-Pastur (MP) distribution [33], which is fully characterized by the noise variance ($\sigma^2$) and data dimensions ($N_q$ and $N_v$). Thus, when $\mathbf{X}$ is noisy data, $P$ and $\sigma$ can be jointly estimated by determining the minimum value of $P$ such that the $\lambda_{P+1}$ to $\lambda_M$ eigenvalues are well described by the MP distribution. A fast and simple approach to do this, introduced by Veraart et al. [32], is to find the minimum $P$ for which the $\left(\lambda_{P+1} - \lambda_M\right) / \left(4\sqrt{(M-P)/N_v}\right) <$ $1/(M-P)\sum_{i=P+1}^{M}\lambda_i$. This works because the inclusion of signal-derived eigenvalues causes the range of $\lambda$ (left hand side of the test inequality) to grow much faster than its sample mean (right hand side). Finally, the denoised signal can be reconstructed using only the $S_{1,1}$ to $S_{P,P}$ singular values, corresponding singular vectors $\mathbf{U}(:, 1:P)$ and $\mathbf{V}(:, 1:P)$, and $\bar{\mathbf{x}}$.

For the mouse brain images, an intensity threshold was used to define a mask of the whole brain, and within this mask, the denoising was performed over a $N_w \times N_w$ moving window. For the simulated images, the mask included the entire $100 \times 100$ image space. In both cases, the window size was defined as $N_w = \text{ceil}\left[\sqrt{N_E}\right]$; a smaller $N_w$ would result in fewer principal components for each moving window step, while a larger $N_w$ would increase the input signal variation within the window without providing more principal components. However, as discussed below, the optimal strategy for defining the size, shape, or other definition of the region used as input for each denoising step remains an open question. Also, while this denoising algorithm, including the joint estimation of $P$ and $\sigma$, is equally valid for real or complex signals, it is not, strictly speaking applicable to magnitude MRI data. (To be clear, the MP-distribution is twice as large for complex compared to real noise, but this scales the range and mean identically, so the test inequality remains the same.) For Rice- or non-central-$\chi$-distributed magnitude data, which may be the only data available from an MRI study, the $\lambda_{P+1}$ to $\lambda_M$ eigenvalues are not MP-distributed. Nonetheless, the same denoising algorithm is applied here to both the complex and magnitude data, and the performance difference between the two scenarios is compared.

## 2.3.2 | Parameter Estimation

From each voxel of both the noisy and denoised MSE image sets, $T_2$ was estimated as a spectrum using the freely available multi-exponential relaxation analysis (MERA) toolbox for MATLAB. Briefly, echo magnitudes were fitted in a non-negative least squares sense with spectrum defined at 100 $T_2$ values, logarithmically spaced between $3/4T_E$ and $4/3N_E T_E$, plus an offset term. For mouse brain data, the analysis included fitting the refocussing flip angle [34] and a constant (across voxels) minimum curvature constraint to smooth each spectrum. Because the simulated data were known to be derived from two discrete $T_2$ components and perfect refocussing, their fitted spectra were not regularized and there was no need to include flip angle estimation. From each $T_2$-spectrum, the estimated short-lived $T_2$ signal fraction ($\hat{f}_s$) and geometric mean $T_2$ of the long-lived signal ($\hat{T}_{2l}$) were computed. The short $T_2$ domain was defined as $T_E < T_2 < 35$ ms for the simulations and $T_E < T_2 < 27$ ms for the mouse brain data. The long $T_2$ domain was everything above the upper boundary of the short $T_2$ domain. Different domains were needed for simulations and experimental data because simulations mimicked relaxation at $3.0$ T while the experimental data were collected at $7.0$ T.

For qMT analysis, the $N_i$ image magnitudes were fitted voxel-wise to the Bloch-McConnell equations describing coupled longitudinal relaxation of water and macromolecular protons [14, 35]. The five fitted model parameters included equilibrium magnetizations of the macromolecular ($M_{0m}$) and water ($M_{0w}$) proton pools, the rate constant of magnetization transfer from the macromolecular to water pool ($k_{mf}$), the longitudinal relaxation rate of the water protons ($R_{1f}$), and the efficiency of the inversion pulse on the water magnetization ($S_w$). The corresponding $R_{1m}$ and $S_m$ values were constrained to $1$ s$^{-1}$ and $0.83$, respectively, in accord with prior studies [14]. The macromolecular proton fraction, was defined as $\hat{f}_m = M_{0m}/(M_{0m} + M_{0w})$.

# 3 | RESULTS

## 3.1 | Simulations

For one example scenario ($N_E = 40$, $T_E = 8$ ms, $\sigma_I = 5$ ms, and SNR = 200), Fig 2 shows original, complex denoised, and difference images. Similar images resulting from magnitude denoising can be seen in supporting Fig S1. Because $T_{2s}$ and $T_{2l}$ varied randomly across the image, the effect of the denoising is difficult to appreciate from the images, but it is easy to see that the sharp boundaries of regions with non-zero $f_s$ were retained. The difference images are somewhat more informative, showing little or no structure, with the exception of some features in the lower part of the difference images at $t_e = 16$ ms. (These second-echo images were chosen as examples because they showed the largest deviations from randomness in the difference image.) These non-random difference image features indicate that, in this region of the image, too few principal components were retained to fully characterize the signal. This is apparent from the maps of the average $P$ value resulting from the MP-PCA denoising for the example simulated images, shown in Fig 4.

For a quantitative assessment of the denoising, the standard deviation of the noise in the denoised images was computed in two ways. A direct calculation was made from the complex difference between the ground truth and denoised images as

$$\hat{\sigma}_d \triangleq \sqrt{\mathrm{var}\left[I_d - I\right]/2},\qquad(3)$$

where var $[\cdot]$ indicates the variance computed across all pixels, and $I_d$ is the denoised image. An approximate estimate of $\hat{\sigma}_d$ was also made using the complex difference between the noisy and denoised images,

$$\hat{\sigma}_{d,approx} \triangleq \sqrt{\left(\sigma^2 - \mathrm{var}\left[I_d - I_n\right]/2\right)}.\qquad(4)$$

This calculation used the known noise variance of the original image ($\sigma^2$), which can be replaced by an empirical estimate when dealing with experimental MRI data (as done below, for the mouse brain images). A plot of these two noise measures from the complex denoised images for the example scenario is shown in Fig 3. The $\hat{\sigma}_d$ values indicate that denoising reduced the standard deviation (SD) of the noise by $\approx 2.5\times$ for the first echo image. The observed decrease in $\hat{\sigma}_d$ with increasing echo time likely reflects some inaccuracies imparted by the denoising. That is, if the denoising does not perfectly capture all the true signal decay, then $\hat{\sigma}_d$ will incorporate both the thermal noise and denoising errors. As the signal amplitude drops with increasing echo time, so will the denoising error. This figure also shows that $\hat{\sigma}_{d,approx}$ serves as a reasonable approximation for $\hat{\sigma}_d$, without the need for a ground truth image data.

Although denoising increased image SNR substantially, of more importance to relaxometry are the effects of denoising on parameter estimates. For the same example scenario as above and five image SNR values, maps of $\hat{f}_s$ are shown in Fig 5. In all cases, it is immediately apparent that denoising either the complex or magnitude images before relaxometry analysis resulted in much smoother $\hat{f}_s$ maps, without apparent loss of resolution. Comparable maps of $\hat{T}_{2l}$ are not particularly informative because the values of $T_{2l}$ varied randomly across the image; however, further quantitative analysis of both $\hat{f}_s$ and $\hat{T}_{2l}$ are shown in Fig 6 and supporting Fig S2 and S3. Scatter plots $\hat{f}_s$ and $\hat{T}_{2l}$ vs ground truth values for example scenarios are shown in supporting Figs S2 and S3. Across all SNR values, the increased precision of both $\hat{T}_{2l}$ and $\hat{f}_s$ is readily apparent. At low SNR, due to the Rician nature of the noise, fitting of the original data resulted in bias of both $\hat{T}_{2l}$ and $\hat{f}_s$. Complex denoising reduced these parameter biases by lowering the image noise floor prior to analysis; magnitude denoising was less effective at removing these biases, because the denoising algorithm retained principal components that characterized the Rician signal offset.

In addition to these Rice-induced biases, at low image SNR, the thin strips of low $f_s$ were not captured in any of the $\hat{f}_s$ maps (see the lower left regions of the maps in Fig 5). This is the same image region that showed structure in the difference images in Fig 2, and so the implication is that the eigenvalues of the principal components needed to capture these small low amplitude fast-relaxing signals were not distinguishable from noise. Nonetheless, in this region, and across the entire image, denoising resulting in lower root mean square error (RMSE) in $\hat{f}_s$. Fig 6 shows RMSE of both $\hat{f}_s$ and $\hat{T}_{2l}$ vs image SNR for the same example scenario ($N_E = 40$, $T_E = 8$ ms, $\sigma_l = 5$ ms). Across all values of image SNR tested, the RMSE of $\hat{f}_s$ dropped by factors of 2.2 to 3.7 due to complex denoising and 1.3 to 3.4 from magnitude denoising. For $\hat{T}_{2l}$, the benefit of denoising was less; RMSE dropped by factors of 1.2 to 2.6 and 1.02 to 2.3 for complex and magnitude denoising, respectively. Importantly, in no case scenario tested, (i.e., all $N_E$, $T_E$ combinations and all values of $\sigma_l$), did the RMSE of $\hat{f}_s$ increase due to denoising. Across all scenarios, the factors of RMSE reduction for $\hat{f}_s$ ranged 1.1 to 4.9 and 1.1 to 3.8 for complex and magnitude denoising, respectively. The corresponding ranges for $\hat{T}_{2l}$ were 0.9 to 6.3 and 0.9 to 2.3, indicating that there were cases where denoising resulted in greater RMSE. However, this only happened in a few high precision cases (i.e., high SNR & high $N_E$) where the RMSE of $\hat{T}_{2l} \ll 1$ ms, even without denoising. Supporting Fig S4 shows RMSE reduction factors for some additional example scenarios. Briefly, these indicate that i) denoising provides a greater benefit when there is less inter-voxel variation in relaxation rate (i.e., smaller $\sigma_l$), and ii) that the dependence on $N_E$ is similar to the dependence on SNR, as one might expect.

## | MRI

Example mouse brain images (original, complex denoised, and the difference images) are shown in Fig 7 for MSE ($N_A = 12$) and IR ($N_A = 14$) acquisitions, at three different contrast levels each. The impact of denoising on image quality is visually apparent and the difference images show no apparent structure, suggesting that the MP-PCA primarily removed noise. The image SNR was calculated as SNR = $\overline{WM}/\hat{\sigma}$, where $\overline{WM}$ was the mean of a white matter region at the first echo/inversion time. The SD of the noise was calculated as $\hat{\sigma} = \overline{BK}/\sqrt{\pi/2}$, where $\overline{BK}$ was the mean of a background region. For the original images in Fig 7, SNRs were 128 (MSE) and 65 (IR), which are roughly the lowest SNR values that one would use for quantitative MET2 and qMT parameter mapping, respectively. The factor of SNR increase that resulted from denoising was estimated as $\hat{\sigma}/\hat{\sigma}_{d,approx}$ (see Eq (4)). For the example images in Fig 7, these factors ranged from $1.4\times$ to $2.3\times$. Across all images (echo/inversion times and $N_A$), the mean image SNR increase factor was $1.9\times$ for the MSE data and $1.8\times$ for the IR data. A plot of all image SNR increases can be found in supporting Fig S5, which shows some variation with $t_e$ or $t_i$ and little/no variation with $N_A$/SNR over the domain tested.

Looking more specifically at the impact of denoising on parameter estimation, Figs 8 and 9 show example $\hat{f}_s$ and $\hat{f}_m$ maps from one mouse brain at five different image SNR values. In both cases, the top row shows the parameter maps resulting from analysis of the original data and the third row shows corresponding maps derived from complex denoised images. In lieu of ground truth, parameter maps from original data at the highest image SNR were used as the reference. Thus, the second row shows the difference between original parameter maps and this reference, and the fourth row so the corresponding differences for the maps from the denoised images. For both MET2 and qMT analyses, denoising the image data improved the resulting parameter map quality, as demonstrated in the first and third rows. In particular, the sharp details of the white matter structures were retained. The $\hat{f}_m$ difference maps show almost no structure, indicating that the denoising imparted little or no bias or spatial artifact on the qMT analysis. For the $\hat{f}_s$ maps, the difference images do show structure, particularly in the external capsules and at lower image SNR.

However, the biased $\hat{f}_s$ values were present in maps from both the original and denoised images, indicating these errors were not a consequence of the denoising procedure.

Across three different mouse brains and three somewhat different imaging slice locations, the impact of denoising on parameter maps was similar, as can be seen in supporting Fig S6. Also, example parameter maps resulting from complex and magnitude denoised images are shown in supporting Figs S7 and S8. At the low end of the SNRs tested, there are some differences in the maps derived from complex vs magnitude denoising, but for moderate levels of SNR and above, the results were effectively the same, as expected.

## DISCUSSION

The simulations and mouse brain MRI demonstrate that MP-PCA denoising can provide substantial improvement to the precision of parameters estimated from multi-exponential MRI signal models. Simulations of MET2 relaxometry meant to mimic myelin water imaging at 3.0 T showed that RMSE of the fast relaxing signal fraction—i.e., the MWF— was reduced by $2 - 4\times$ for typical imaging conditions. Experimental studies in mouse brains at 7.0 T showed roughly similar precision improvements for MWF maps and macromolecular pool fraction measured by qMT. For example, Fig 8 shows comparable quality $\hat{f}_s$ maps derived from denoised data at SNR = 128 to $\hat{f}_s$ maps derived from original data at SNR = 367. Likewise, in Fig 9, $\hat{f}_m$ maps derived from denoised data at SNR = 87 are superior to $\hat{f}_m$ maps derived from original data at SNR = 138. Also, in both the simulations and experimental studies presented here, these parameter precision gains come without an apparent resolution cost or bias. Of course, there are a number of issues to be considered and/or studied further.

For any PCA/SVD denoising or model reduction, choosing the number of components to retain can be a challenge. Too much rank reduction will result in a model that cannot adequately describe the underlying data, and important details can remain lost in the noise. Figure 10 shows plots of the PCA eigenvalues from an example $N_v$ voxels overlapping the corpus callosum near the middle of four different mouse brain image sets. For reference, also plotted in each frame are the eigenvalues resulting from randomly generated noise with standard deviation matching that estimated from the image background. The $P$ eigenvalues retained by the MP-PCA algorithm are identified in each frame, and it it reasonable to surmise that various algorithms could be used to identify these same values. However, the Marchenko-Pastur [33] distribution offers a theoretically attractive approach for choosing the signal rank cut-off, and because it is data driven, no a priori assumptions about the underlying signal model are needed. In the present implementation, the joint estimation of $P$ and $\sigma$ is computationally fast, permitting denoising of $N_x \times N_y \times N_E = 100 \times 100 \times 40$ complex images in $\approx 5$ sec using MATLAB on desktop computer.

This study has not attempted to modify or optimize the MP-PCA image denoising algorithm, but rather has simply applied the method as previously presented [32], with the minor modification of removing the mean signal across voxels prior to SVD factorization. With this algorithm, each PCA denoising step is applied to a local square (or cube, for 3D imaging) of $N_v$ voxels, which means that the denoising performance will vary with tissue heterogeneity over local region. For accurate parameter estimation, the eigenvalues of a sufficient number of principal components must be well distinguished from the noise in order to characterize the relaxation of all $N_v$ voxels. We have reasoned that $N_v \approx N_q$ offers the benefit making full use of the $N_q$ images while minimizing the number of components likely needed to characterize the variation of relaxation across the $N_v$ voxels, but this has not been rigorously examined. Further, it is possible that a non-local approach, where the $N_v$ voxels are selected based on similar relaxation characteristics, will offer performance advantages.

An intuitive explanation for the efficacy of the MP-PCA denoising on multi-exponential relaxometry stems from the ill-conditioned nature of the problem. Consider a linear system, $\mathbf{y} = \mathbf{As}$, where $\mathbf{y}$ is a column of echo magnitudes, **s** is the spectrum of signal amplitudes as a function of $T_2$ (i.e., the $T_2$-spectrum), and the columns of **A** are decaying exponential functions spanning the relevant $T_2$ domain. For the case of $N_E = 40$, $T_E = 8$ ms, and **A** comprised of 100 columns of exponential decays spanning $T_2 = 10$ ms to 500 ms, the rank of **A** is equal to 18 at double precision round-off tolerance and only 8 using a tolerance of $1 \times 10^{-3}$. This is to say that any linear combination of the columns of **A** can be expressed effectively (i.e., within the precision of practically any MRI measurement) as a linear combination of only a handful of singular vectors, typically many fewer than the number of echoes acquired. This characteristic has been used for some time to reduce the dimensionality of the numerical inverse Laplace transform [36] (and is part of the aforementioned MERA processing code), and the same idea underlies multi-echo image denoising [23] and sub-sampling strategies for fast parameter mapping [37, 38, 26]. On one hand, this rank deficiency is what makes the inverse solution—estimate of the $T_2$ spectrum—difficult. On the other hand, it means that, even for relatively high image SNR, only a handful of principal components are necessary to characterize all of the signals from a region of voxels each with somewhat different relaxation characteristics. The same is true for a mono-exponential problem, but the practical payoff of denoising is less because the solution is not ill-conditioned, meaning that a relatively precise estimate of $T_2$ can be achieved with relatively low image SNR.

## CONCLUSIONS

The use of MP-PCA denoising offers substantial improvement in the precision of parameters estimated from MRI multi-exponential relaxometry. Of particular interest for neuroimaging, the results indicate that the precision of myelin specific parameter maps may be increased by a factor $\sim 2\times$ or more.

## ACKNOWLEDGEMENTS

The authors acknowledge Dr. Robert Carson and Mr. Dong Kyu Kim for providing the fixed mouse brains used in this study. The MERA toolbox for MATLAB is available at https://github.com/markdoes/MERA and the MP-PCA image denoising code is available at XXXXXXXXX.

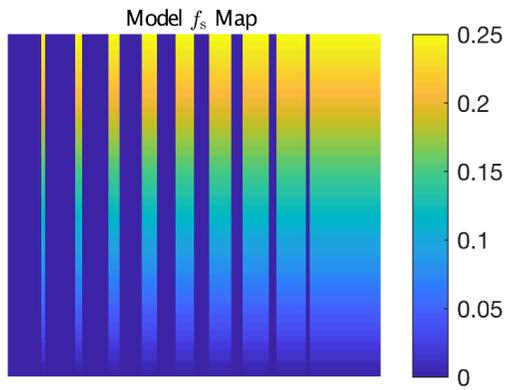

**FIGURE 1**   Model map of $f_s$ used for the simulations. The non-zero values of $f_s$ ranged 0 to 0.25 in vertical strips of widths ranging 1 to 9 pixels

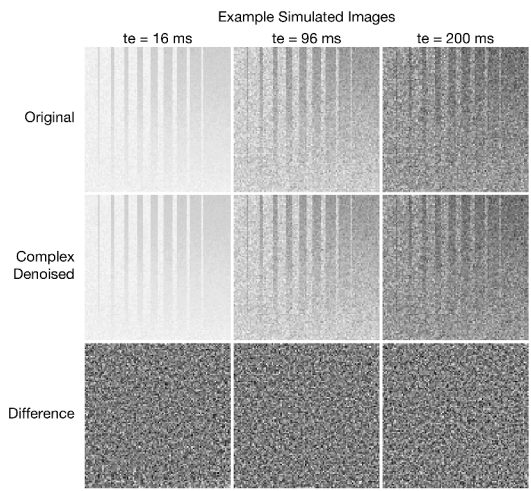

**FIGURE 2** Example simulated images from the scenario with $N_E = 40$, $T_E = 8$ ms, $\sigma_I = 5$ ms, and SNR = 100. The top row shows original noisy images. The second and third rows show the complex denoised images and the difference images, respectively. At each different $t_e$, the original and denoised images are all displayed using the same grayscale, and all three differences images are scaled to $\pm 3\sigma$

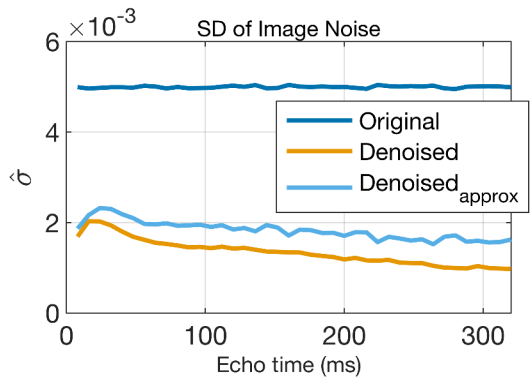

**FIGURE 3** Calculated image SNR values before and after denoising.

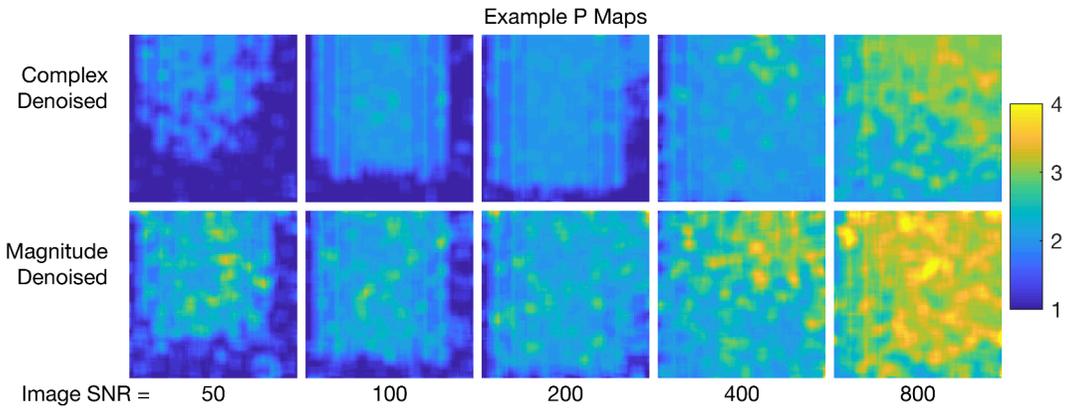

**FIGURE 4**  Maps of the average *P* value (number of retained principal components in the MP-PCA denoising) corresponding to the example images in Fig 2.

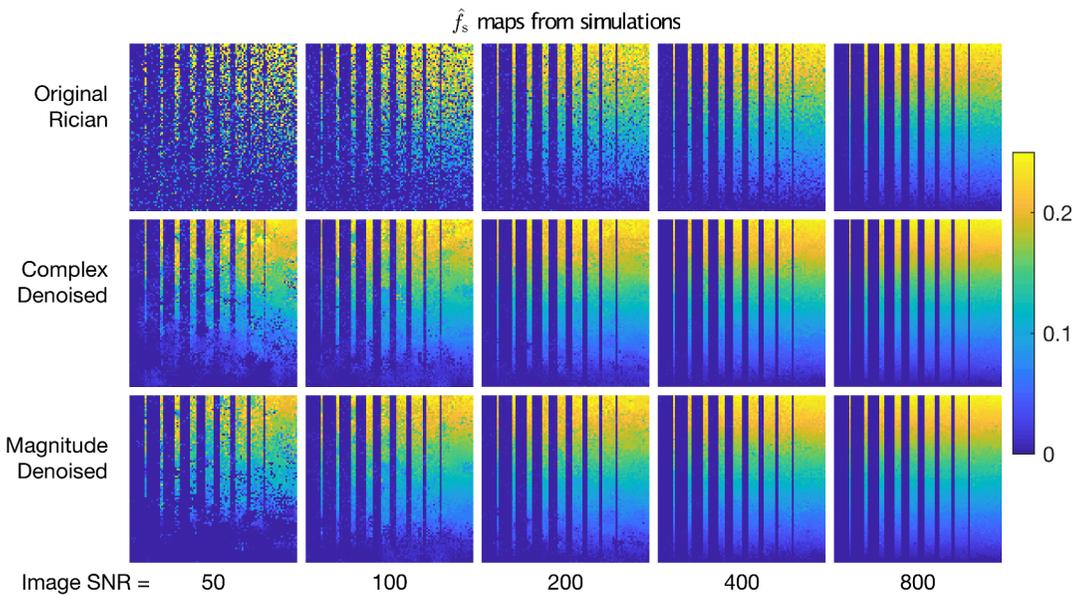

**FIGURE 5**   Maps of estimated short $T_2$ signal fraction from simulated images with five different SNR values. The top row shows results from relaxometry analysis of the original images, while the middle and bottom rows show results from analysis of complex and magnitude denoised images, respectively.

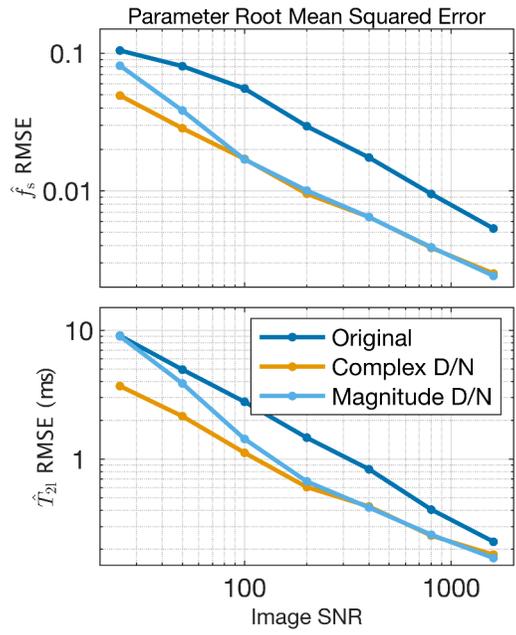

**FIGURE 6** RMSE values of $\hat{f}_s$ and $\hat{T}_{2l}$

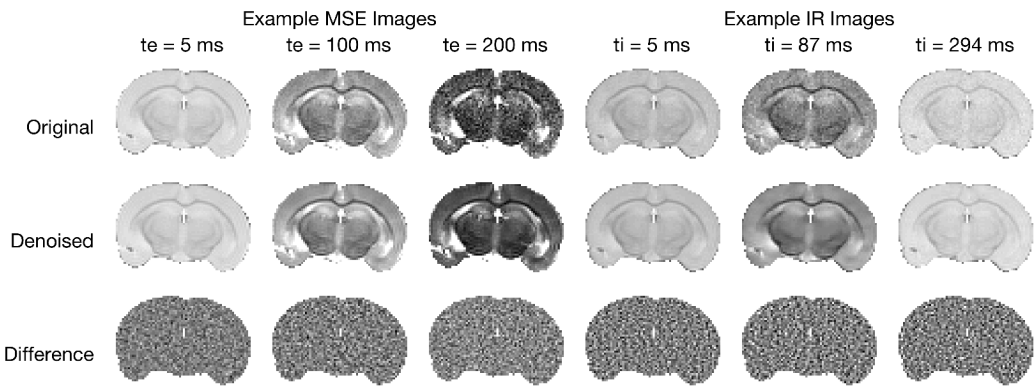

**FIGURE 7** Images from the $N_A$ = 8 acquisitions at two echo times: (left column) $t$ =5 ms and (right column) $t$ =100 ms. The top row shows the raw images, the middle row shows the corresponding denoised images, and the bottom row shows the difference images. The denoised images retain sharp anatomical boundaries, and the difference images show little to no structure.

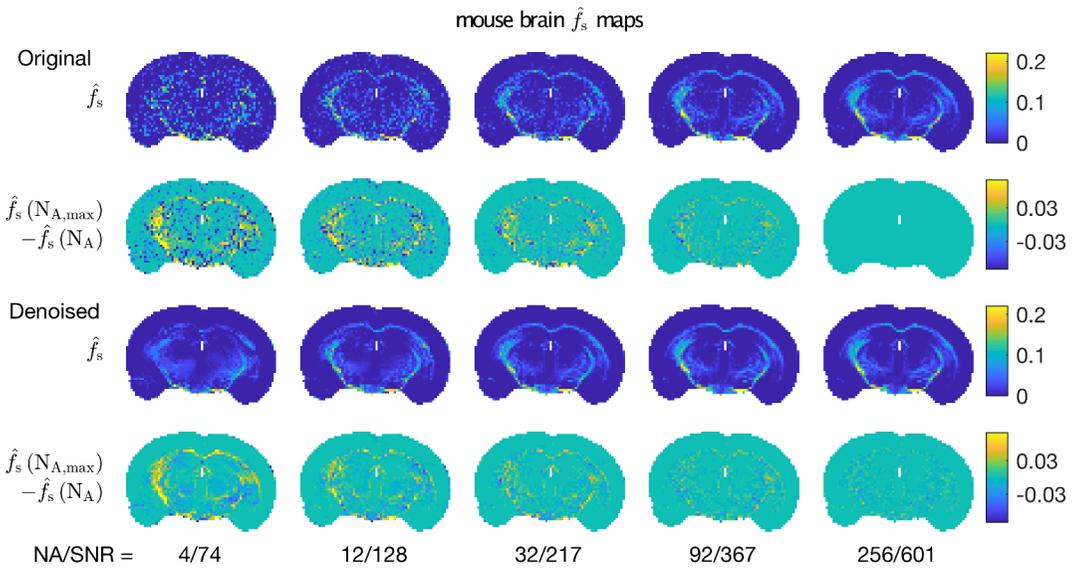

**FIGURE 8** Maps of $\hat{f}_s$ from (top) original and (2nd row) denoised mouse brain images, for five different $N_A$/SNR values. The 3rd row shows differences between these maps, original-denoised.

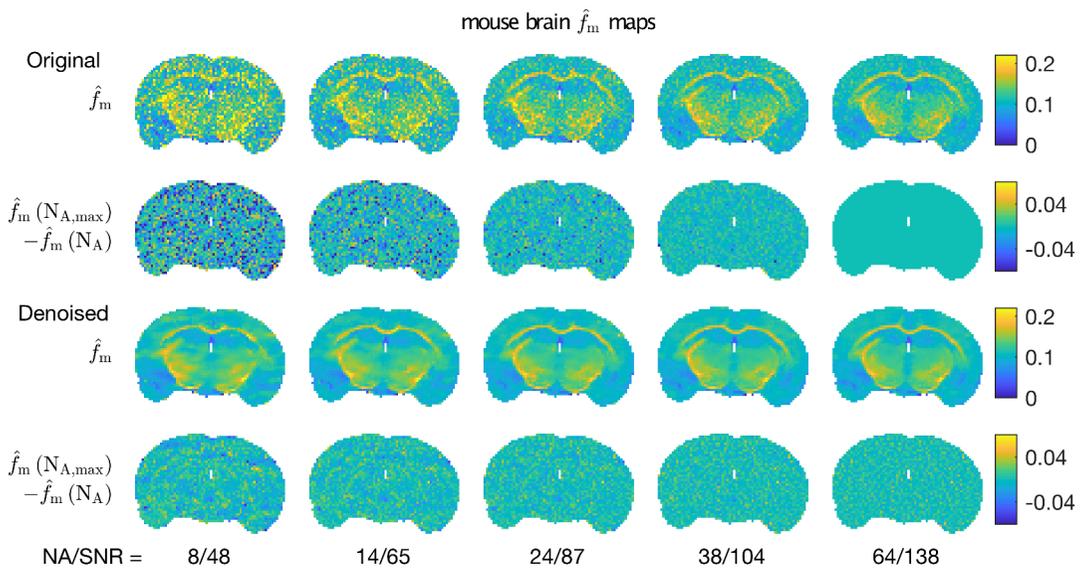

**FIGURE 9**  Maps of $\hat{f}_m$ from (top) original and (2nd row) denoised mouse brain images, for five different $N_A$/SNR values. The 3rd row shows differences between these maps, original-denoised.

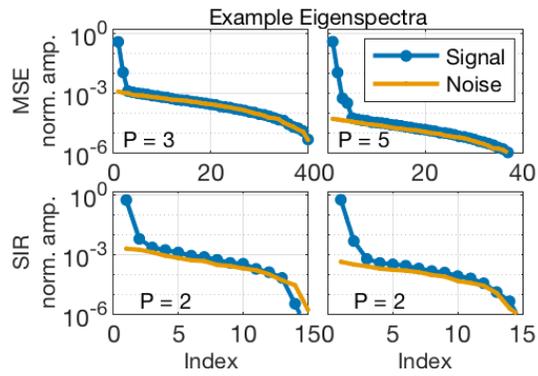

**FIGURE 10** Eigenvalues from the PCA analysis of example $N_v$ voxel regions overlapping the corpus callosum in mouse brain. The top row shows result from MSE images ($N_v = 49$), acquired with $N_A = 12$ (left) and 256 (right). The bottom shows results from SIR images ($N_v = 16$), acquired with $N_A = 3$ (left) and 64 (right).

## | Supporting Figures

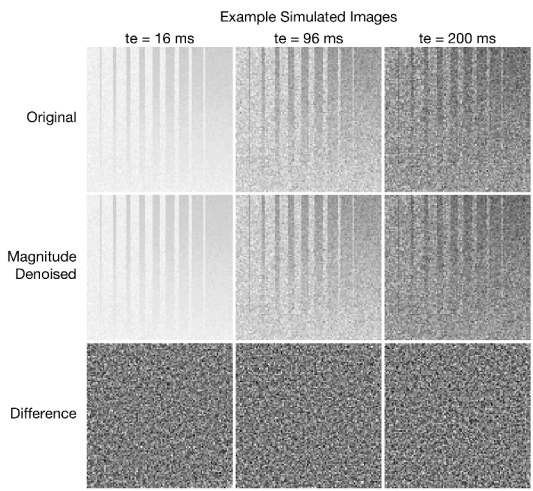

**FIGURE S1** Example simulated images from the scenario with $N_E = 40$, $T_E = 8\,\text{ms}$, $\sigma_I = 5\,\text{ms}$, and SNR = 100. The top row shows original noisy images. The second and third rows show the magnitude denoised images and the difference images, respectively. At each different $t_e$, the original and denoised images are all displayed using the same grayscale, and all three differences images are scaled to $\pm 3\sigma$

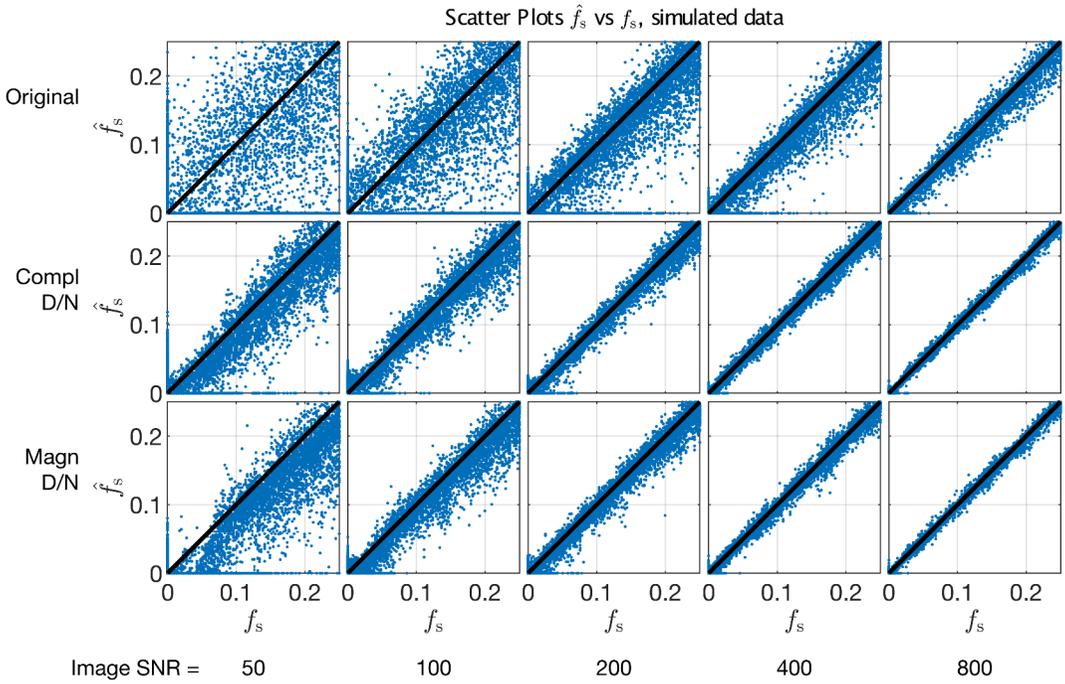

**FIGURE S2** Scatter plots of estimate short $T_2$ signal fraction, $\hat{f}_s$, vs the ground truth, $f_s$, for the example scenario of $N_E = 40$, $T_E = 8$ ms, and $\sigma_I = 5$ ms.

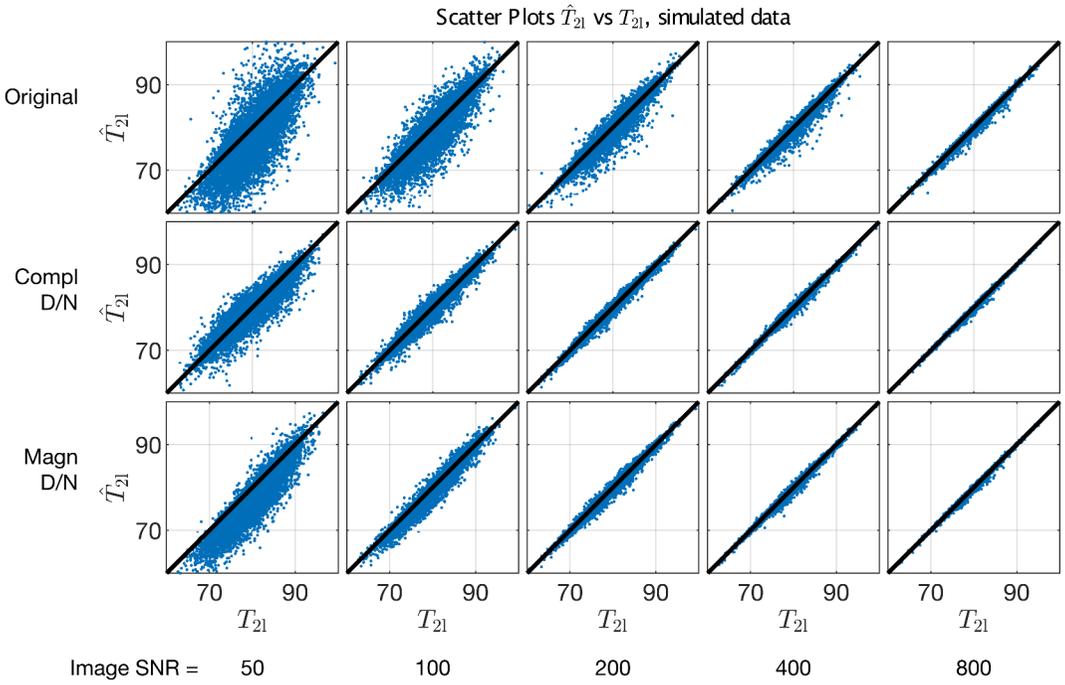

**FIGURE S3** Scatter plots of estimate long $T_2$, $\tilde{T}_{2l}$, vs the ground truth, $T_{2l}$, for the example scenario of $N_E = 40$, $T_E = 8\,\text{ms}$, and $\sigma_I = 5\,\text{ms}$.

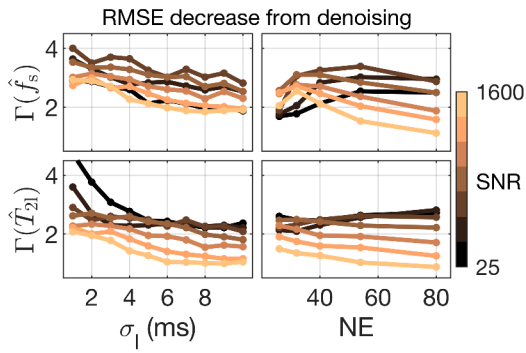

**FIGURE S4** Measure of the relative decrease in parameter RMSE due to denoising, as a function of $\sigma_I$ (left), $N_E$ (right), and image SNR (color). For all frames, the vertical axis is $\Gamma(p) \triangleq \mathrm{RMSE_o}\,(p)\,/\mathrm{RMSE_d}\,(p)$, where subcripts 'o' and 'd' indicate 'original' and 'denoised'.

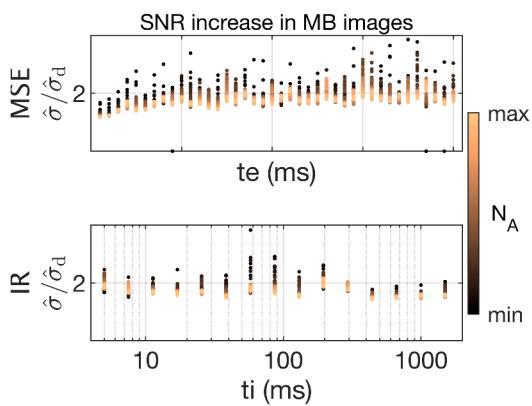

**FIGURE S5**  Relative increase in mouse brain image SNR after denoising, as a function of $t_e$ (MSE, top) or $t_i$ (IR, bottom) and $N_A$ (color). Note that $N_A$ ranges 4 to 256 for MSE and 1 to 64 for IR.

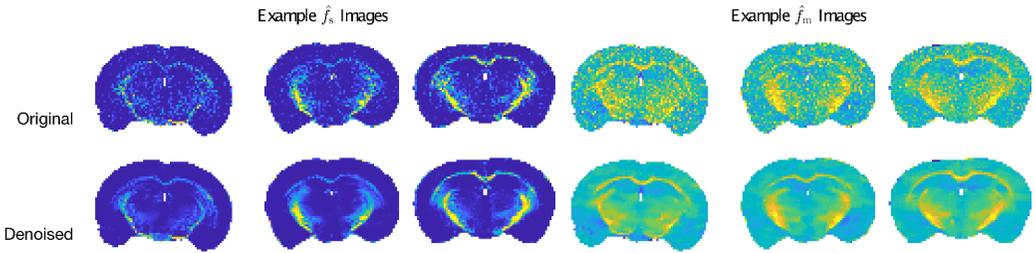

**FIGURE S6** Example $\hat{f}_s$ and $\hat{f}_m$ parameter maps from three different mouse brains. Parameter intensities vary some between brains/slice location, but the improvement in parameter map quality due to denoising is qualitatively similar in all cases.

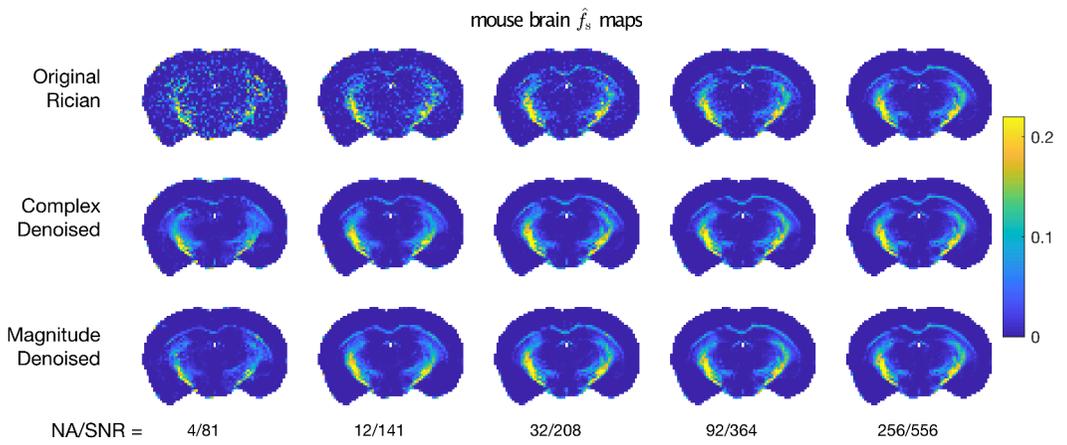

**FIGURE S7** A comparison of $\hat{f}_s$ maps from original, complex denoised, and magnitude denoised images. Except at low image SNR, the effect of complex and magnitude denoising on $\hat{f}_s$ maps was similar

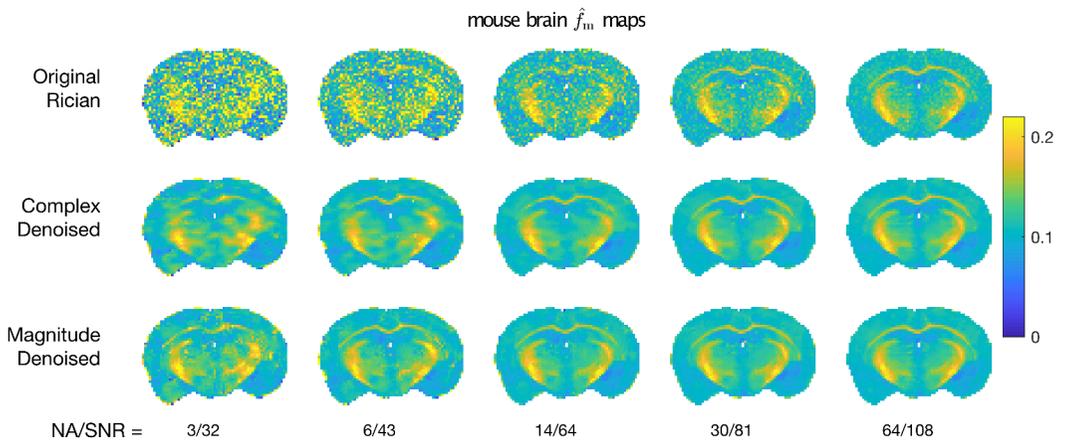

**FIGURE S8** A comparison of $\hat{f}_m$ maps from original, complex denoised, and magnitude denoised images. Except at low image SNR, the effect of complex and magnitude denoising on $\hat{f}_m$ maps was similar